\documentclass[twocolumn,showpacs,preprintnumbers,amsmath,amssymb,prb,aps]{revtex4-1}
\usepackage{epsfig}
\usepackage{epstopdf}
\usepackage{graphicx}
\usepackage{dcolumn}
\usepackage{bm}
\usepackage{slashbox}
\usepackage{textcomp}
\begin{document}

\title{Studying the effect of different exchange correlation functionals on the structural and electronic properties of a half-Heusler NaAuS compound}
\author{Antik Sihi$^1$}
\altaffiliation{Electronic mail: sihiantik10@gmail.com}
\author{Sohan Lal$^2$}
\author{Sudhir K. Pandey$^2$}
\affiliation{$^1$School of Basic Sciences, Indian Institute of Technology Mandi, Kamand 175005, Himachal Pradesh, India\\
       $^2$School of Engineering, Indian Institute of Technology Mandi, Kamand 175005, Himachal Pradesh, India}

\date{\today}

\begin{abstract}
Theoretically, NaAuS is predicted as topological insulator, while no detail electronic structure study has been done for this compound. Here, we report the structural and electronic properties of NaAuS by using LDA, PBEsol, PBE and revPBE exchange correlation functionals. The calculated values of equilibrium lattice constant for LDA, PBEsol, PBE and revPBE exchange correlation functionals are found to be $\sim$6.128 $\AA$, $\sim$6.219 $\AA$, $\sim$6.353 $\AA$ and $\sim$6.442 $\AA$, respectively. The bulk modulus predicted by LDA, PBEsol, PBE and revPBE exchange correlation functionals is $\sim$66.6, $\sim$56.4, $\sim$46.5 and $\sim$39.3 GPa, respectively. Hence, the order of calculated values of bulk modulus is consistent with the order of calculated values of equilibrium lattice parameters for these exchange correlation functionals. The spread of total density of states below the Fermi level decreases as the exchange correlation functional changes from LDA to PBEsol to PBE to revPBE, which is also found to be consistent with the order of bulk modulus for these exchange correlation functionals. In presence of spin-orbit coupling, a direct band gap is observed in NaAuS compound, which is found to be $\sim$0.26, $\sim$0.25, $\sim$0.24 and $\sim$0.23 eV for LDA, PBEsol, PBE and revPBE exchange correlation functionals, respectively. Here, NaAuS is found to be topological insulator as it shows band inversion at $\Gamma$ point. The calculated values of band inversion strength for LDA (PBEsol) and PBE (revPBE) exchange correlation functionals are $\sim$1.58 eV ($\sim$1.57 eV) and $\sim$1.50 eV ($\sim$1.47 eV), respectively.

\end{abstract}


\maketitle

\section{Introduction} 

  Heusler alloys are taken great importance to study their various physical properties after discovered by Fredrick Heusler in 1903 \cite{Starck,Heusler}. Generally, the Heusler alloys are divided into two categories, one is called half-Heusler alloy with chemical formula MM$'$X and another is called full-Heusler alloy with chemical formula M$_{2}$M$'$X. These alloys have special feature that their properties differ completely from those of the contained elements. For example, Cu$_{2}$MnAl (full-Heusler alloy) is ferromagnetic, where Cu and Al atoms are non-magnetic and Mn atom is anti-ferromagnetic by themselves \cite{Geiersbach}. Other example is TiNiSn (half-Heusler alloy), which is  semiconducting, even though it is made of three components that are metals \cite{Aliev}. New trend comes to find the topological insulation property from ternary half-Heusler materials in condensed matter physics \cite{Chadov,Lin,Yan,Wray,Xiao,Li,Muchler}. Half-Heusler compounds are usually nonmagnetic and semiconducting when the number of total valence electrons is 18,
\begin{equation}     
V(M)+V(M')+V(X)=18   \nonumber
\end{equation}                              
 which is known as 18-electron rule \cite{Jung}. This rule in general is used for searching the topological insulators (TIs)  \cite{Shi}. Hence, half-Heusler material takes great attention to predict TIs. First-principles calculations have been widely used to predict topological insulators with great success \cite{Zhang}.
    
    The materials which are insulating in their interior but can support the flow of electrons on their surface, are called TIs. TIs are currently creating a new division of research activities in condensed matter physics \cite{Hasan,Moore,Qi}. First TI was theoretically predicted by Bernevig $et$ $al.$ and experimentally observed by Konig $et$ $al.$ in the HgTe quantum well \cite{Bernevig,Konig}. Bernevig $et$ $al.$ have shown that the band inversion property driven by spin-orbit coupling (SOC) is used to discover the TIs. Sawai $et$ $al.$ have observed that the topology of the electronic band structures can be described by band inversion between $\Gamma_6$ and $\Gamma_8$ energy levels at the $\Gamma$ symmetry point in the Brillouin zone and they define the band inversion strength $\Delta $ as the energy difference between these two states, i.e., ${\Delta=[E_{\Gamma_8}-E_{\Gamma_6}]}$, where the $E_{\Gamma_6}$ and $E_{\Gamma_8}$ are the energy levels for $\Gamma_6$ and $\Gamma_8$ at the $\Gamma$ point \cite {Al}. As compared with an ordinary superconductor, TIs have an important role for creating topological quantum computation \cite{Fu}. In order to find the new topological insulators, Lin $et$ $al.$ have studied theoretically more than 2000 half-Heusler compounds \cite{Shi}. Out of these, only LiAuS and NaAuS compounds have found to be TIs with band gap $\sim$0.20 and $\sim$0.19 eV, respectively. They have studied both compounds using only PBE and hybrid density functionals in face-centered-cubic (FCC) phase. As per our knowledge, the synthesis of NaAuS compound (which we have chosen to study in the present manuscript) in the FCC phase is not available in the literature. However, this compound is synthesized experimentally in orthorhombic phase \cite{Axtell}. In general, half-Heusler compounds are described by space group \textit{F}-43m. Hence, in NaAuS half-Heusler with space group \textit{F}-43m, Na, Au and S atoms are located at Wyckoff positions (1/2, 1/2, 1/2), (1/4, 1/4, 1/4) and (0,0,0), respectively. The detailed structural and electronic study of NaAuS compound is not found in the literature \cite{Shi}. Hence, it will be interesting to see the effect of different exchange correlation functionals on the structural and electronic behaviour of this compound in more detail.
      
      Here in the present work, we study the structural and electronic behaviour of NaAuS compound by using density functional theory. The order of calculated values of equilibrium lattice parameter for LDA$>$PBEsol$>$PBE$>$revPBE exchange correlation functionals, which is consistent with the order of bulk modulus predicted by these exchange correlation functionals. Among these exchange correlation functionals, the total density of states below the Fermi level are spread in more region for LDA and less region for revPBE, which indicates that the bulk modulus predicted by LDA is largest and revPBE is smallest. The order of calculated values of direct band gap is found to be LDA$>$PBEsol$>$PBE$>$revPBE exchange correlation functionals. The band inversion is found at $\Gamma$ point for NaAuS indicates that this compound is a topological insulator. The order of magnitude of calculated values of band inversion strength is found to LDA$>$PBEsol$>$PBE$>$revPBE exchange correlation functionals.

\section{Computational details}
   
    The electronic structure calculation of NaAuS compound is performed by using the full-potential linearized-augmented plane-wave (FP-LAPW) method as implemented in Elk code \cite{elk}. Here we have employed LDA, PBEsol, PBE and revPBE exchange correlation functionals \cite{Perdew, Pbesol, GGA, Yang}. SOC is also considered to see its effect on the electronic properties of this compound. The muffin-tin sphere radii used for Na, Au and S atoms are 2.0, 2.5 and 2.0 bohr, respectively. The values of \textit{rgkmax} and \textit{gmaxvr} are set to be 8 and 14, respectively. These values are sufficient to get nice parabolic energy versus volume curves. $10\times 10\times 10$ k-point mesh size has been used in the present calculations. Total energy to reach convergence has been set below $10^{-4}$ Hartree/cell.

  The equilibrium lattice parameters are computed by fitting the total energy versus unit cell volume data to the universal equation of state \cite{Vinet}. The universal equation of state is defined as,

 \begin{center}{$P$ = [3$B$$_{0}$(1 - $\chi$)/$\chi$$^{2}$]e$^{3/2(B'_{0}-1)(1-\chi)}$,  $P$ = -($\partial$$E$/$\partial$$V$)}\end{center}   

       where $P$, $E$, $V$, $B$$_{0}$ and $B$$_{0}$$^{'}$ are the pressure, energy, volume, bulk modulus and pressure derivative of bulk modulus, respectively and $\chi$ = ($V$/$V_{0}$)$^{1/3}$. 
       
\begin{table}[ht]
\caption{Calculated equilibrium lattice constant (\AA), bulk modulus (GPa), band gap (eV) and band inversion strength (eV) of NaAuS compound (where calculated values in presence of spin-orbit coupling are shown by bold).\\}
\centering 
\begin{tabular}{p{2.0cm}p{1.5cm}p{1.4cm}p{1.2cm}p{1.5cm}}
\hline
\hline
{Exchange correlation functionals}&{Lattice constant (‎\AA)}&{Bulk modulus (GPa)}&{Band Gap (eV)}&{Band inversion strength (eV)}\\[0.0ex]
&&\\[0.0ex]
\hline
LDA&6.128 (\bf 6.107)&66.6 (\bf 69.0)&\bf 0.26&\bf 1.58\\
PBEsol&6.219 (\bf 6.196)&56.4 (\bf 58.7)&\bf 0.25&\bf 1.57\\
PBE&6.353 (\bf 6.328)&46.5 (\bf 48.4)&\bf 0.24&\bf 1.50\\
revPBE&6.442 (\bf 6.416)&39.3 (\bf 41.1)&\bf 0.23&\bf 1.47\\[1ex]
\hline
\hline
\end{tabular}
\label{table:the exp}
\end{table} 
    
\section{Results and Discussion}
  
    The total energy difference between the energies (function of volume) and energy corresponding to the equilibrium volume [$\Delta$$E$=$E$(V)-$E$(V$_{\rm eq}$)] per formula unit versus primitive unit cell volume of NaAuS for LDA, PBEsol, PBE and revPBE exchange correlation functionals have been plotted in Fig. 1. A parabolic behavior is observed in each curves for every exchange correlation functional. It is also clear from the figure that the equilibrium volume is shifted towards the higher value as the exchange correlation functional changes from LDA $\rightarrow$  PBEsol $\rightarrow$  PBE $\rightarrow$ revPBE. In order to determine the equilibrium volume and bulk modulus for the different exchange correlation functional, we have fitted the data of the total energy-volume by using the universal equation of state. The equilibrium volume corresponding to minimum energy predicted by LDA, PBEsol, PBE and revPBE exchange correlation functionals are $\sim$388.2, $\sim$405.7, $\sim$432.7 and $\sim$450.9 bohr$^{3}$, respectively. The equilibrium lattice constant for different exchange correlation functionals is obtained from the equilibrium volumes of respective exchange correlation functional. The calculated values of equilibrium lattice constant (bulk modulus) of NaAuS for LDA, PBEsol, PBE and revPBE are $\sim$6.128 $\AA$ ($\sim$66.6 GPa), $\sim$6.219 $\AA$ ($\sim$56.4 GPa), $\sim$6.353 $\AA$ ($\sim$46.5 GPa) and $\sim$6.442 $\AA$ ($\sim$39.3 GPa), respectively. These values of equilibrium lattice constant and bulk modulus are shown in Table 1. It is clear from the table that bulk modulus (equilibrium lattice constant) decreases (increases) as exchange correlation functional changes from LDA to PBEsol to PBE to revPBE. The reason of opposite behaviour of bulk modulus as compared to equilibrium lattice constant for these exchange correlation functionals will be discussed after the SOC discussion.
    
\begin{figure}
\caption{Total energy difference between the volume dependent energies and energy corresponding to the equilibrium volume [$\Delta$$E$=$E$(V)-$E$(V$_{\rm eq}$)] per formula unit versus primitive unit cell volume plots for (a) LDA, (b) PBEsol, (c) PBE and (d) revPBE exchange correlation functionals without spin-orbit coupling.\\
\\}
\includegraphics[width=1.0\linewidth, height=9cm]{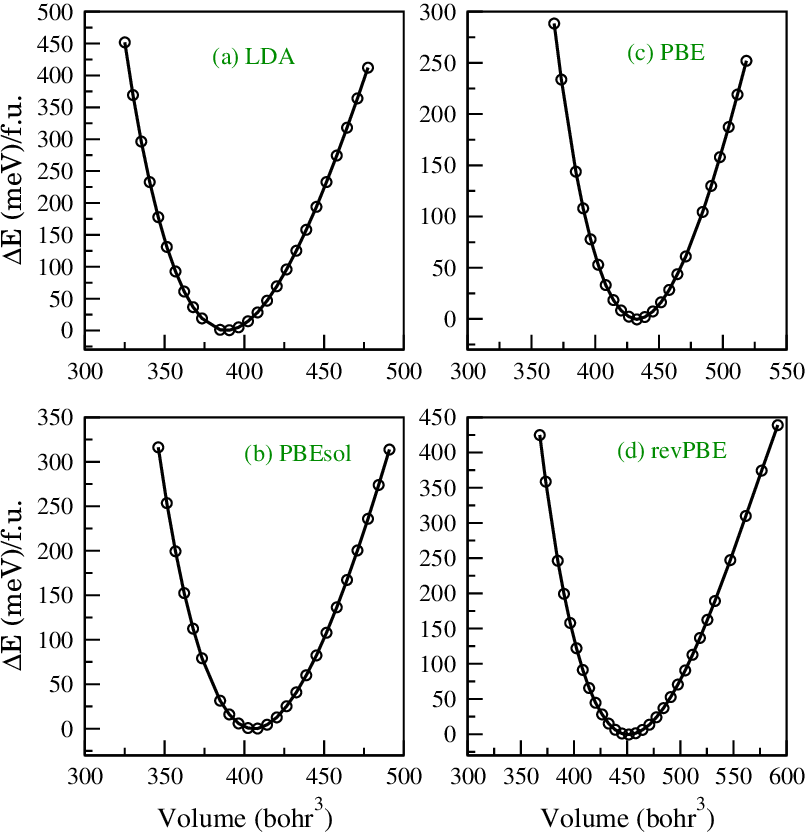}
\end{figure}
      
\begin{figure}
\caption{Total energy difference between the volume dependent energies and energy corresponding to the equilibrium volume [$\Delta$$E$=$E$(V)-$E$(V$_{\rm eq}$)] per formula unit versus primitive unit cell volume plots for LDA, PBEsol, PBE and revPBE exchange correlation functionals with spin-orbit coupling.\\
\\}
\includegraphics[width=1.0\linewidth, height=8.0cm]{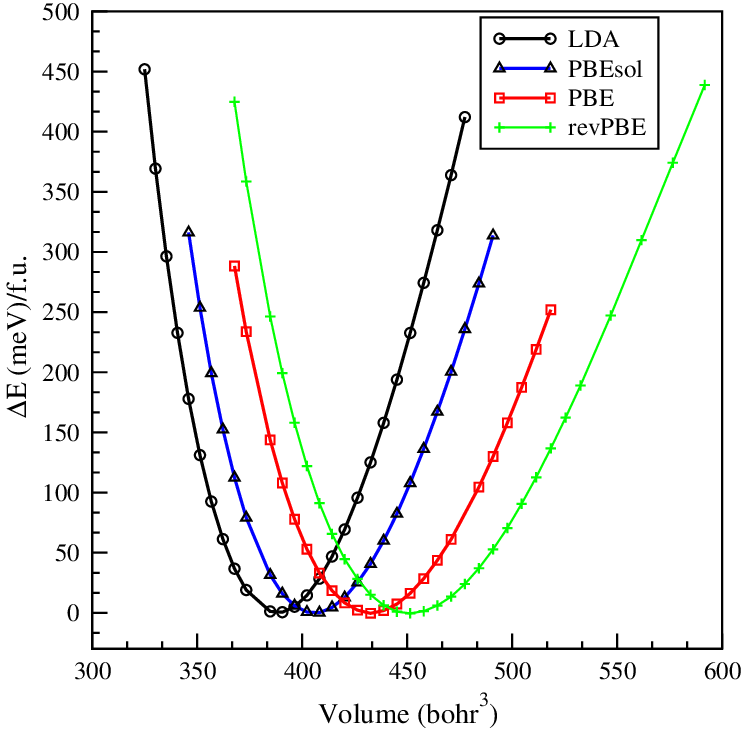}
\end{figure}

      The SOC is expected to play an important role in NaAuS compound because of the heavier Au atom. Hence, we have included the SOC in our calculations. The plot of $\Delta$$E$ per formula unit versus primitive unit cell volume is shown in Fig. 2. It is clear from the figure that a parabolic behavior is also seen here for these exchange correlation functionals, which is similar to that observed without SOC. The equilibrium volume and bulk modulus of NaAuS compound for every exchange correlation functional are also computed by fitting total energy-volume data using the universal equation of state. The equilibrium volume corresponding to minimum energy predicted by LDA, PBEsol, PBE and revPBE exchange correlation functionals are $\sim$384.3, $\sim$401.2, $\sim$427.5 and $\sim$445.6 bohr$^{3}$, respectively. Hence, by including SOC, a small decrease in equilibrium volumes is observed for every exchange correlation functional. The calculated values of equilibrium lattice constant (bulk modulus) of NaAuS for LDA, PBEsol, PBE and revPBE are $\sim$6.107 $\AA$ ($\sim$69.0 GPa), $\sim$6.196 $\AA$ ($\sim$58.7 GPa), $\sim$6.328 $\AA$ ($\sim$48.4 GPa) and $\sim$6.416 $\AA$ ($\sim$41.1 GPa), respectively. These values of equilibrium lattice constant and bulk modulus are also shown in Table 1. It is clear from the table that the values of equilibrium lattice constant and bulk modulus by including SOC does not differ too much from without SOC values for every exchange correlation functional. The values of equilibrium lattice constant differ only about 0.3$\sim$0.4\% for both calculation for every exchange correlation functional, which indicates that SOC does not show any important role to calculate equilibrium lattice constant for NaAuS. However, by including SOC, the values of bulk modulus is about 3.5$\sim$4.5\% greater than without SOC values for every exchange correlation functional. It is also clear from the table that among these exchange correlation functionals, LDA (revPBE) gives the lowest (highest) value of equilibrium lattice constant and highest (lowest) value of the bulk modulus. Generally, LDA overestimates, PBE underestimates and PBEsol slightly overestimates the bulk modulus \cite{Csonka,Sharma}.
   
\begin{figure}
\caption{Total energy difference between the volume dependent energies and energy corresponding to the equilibrium volume [$\Delta$$E$=$E$(V)-$E$(V$_{\rm eq}$)] per formula unit versus total primitive unit cell volume difference between the primitive unit cell volume and the corresponding equilibrium volume [$\Delta$$V$=$V$-V$_{\rm eq}$] for LDA, PBEsol, PBE and revPBE exchange correlation functionals without spin-orbit coupling.\\
\\}
\includegraphics[width=1.0\linewidth, height=7cm]{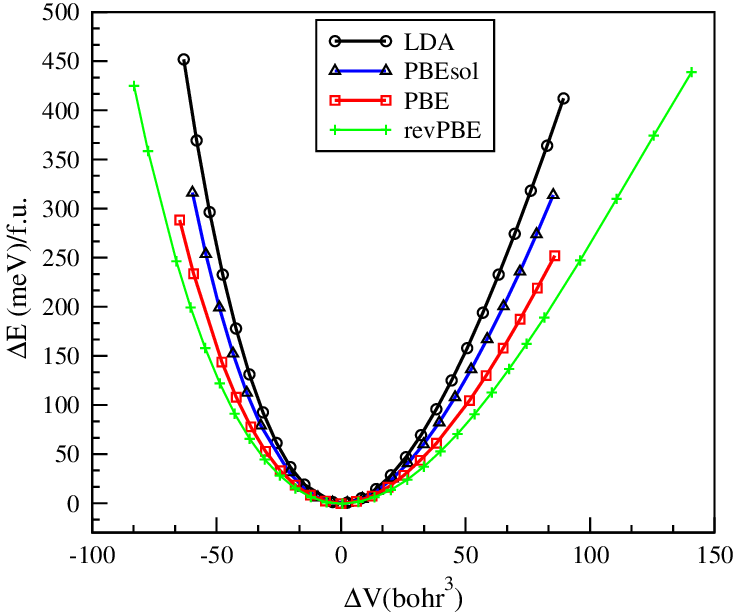}
\end{figure}

    Now, we discuss the cause of opposite behavior of bulk modulus as compared to equilibrium lattice constant, when exchange correlation functional changes from LDA to PBEsol to PBE to revPBE. As we know that the pressure is defined as the rate of change of energy with respect to volume and the bulk modulus is directly proportional to the rate of change of pressure with respect to volume. We have plotted $\Delta$$E$ per formula unit versus total primitive unit cell volume difference between the primitive unit cell volume and the corresponding equilibrium volume [$\Delta$$V$=$V$-V$_{\rm eq}$] for NaAuS using every exchange correlation functional in Fig. 3. It is evident from the figure that LDA (revPBE) gives the most (least) steeper slope, which indicates that LDA (revPBE) gives higher (lower) rate of change of energy with respect to volume. Hence, LDA (revPBE) gives higher (lower) bulk modulus as shown in Table 1. This is due to the direct relationship between energy and bulk modulus.

\begin{figure}
\caption{Total density of states (TDOS) plots for NaAuS compound. (a) TDOS plot using LDA and PBEsol, (b) TDOS plot using PBE and revPBE. Zero energy corresponds to Fermi level.\\
\\}
\includegraphics[width=1.0\linewidth, height=7.0cm]{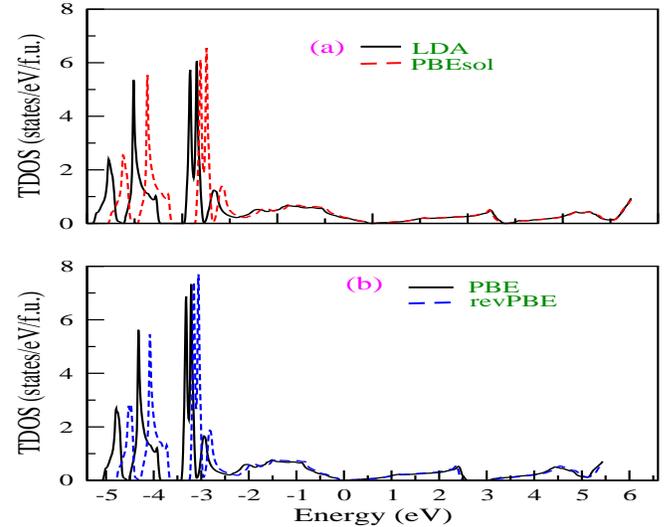}
\end{figure}  
  
    The plot of the total density of states (TDOS) of NaAuS for every exchange correlation functional is shown in the Fig. 4. It is clear from the figure that the every exchange correlation functional gives almost similar behaviour of the TDOS, when SOC is excluded in the calculations. However, a very small TDOS at the Fermi level is obtained for every exchange correlation functional, which indicates a soft band gap. Here, it is important to note that in the valance band (VB), the peaks of TDOS shifted towards the Fermi energy as the exchange correlation functional changes from LDA to PBEsol to PBE to revPBE. Below Fermi level, the TDOS is spread upto -5.9 eV (-5.6 eV) and -5.1 eV (-4.8 eV) for LDA (PBEsol) and PBE (revPBE), respectively. It is interesting to note that the order of bulk modulus and the spread of TDOS below Fermi level is similar for these exchange correlation functional. Hence, in general one can predict the order of bulk modulus by looking the TDOS of these exchange correlation functionals.  
    
\begin{figure}
\caption{Partial density of states (PDOS) plots of Na, Au and S atoms of NaAuS compound using PBEsol exchange correlation functional with and without spin-orbit coupling included in calculations. Zero energy corresponds to Fermi level.\\
\\}
\includegraphics[width=1.0\linewidth, height=9cm]{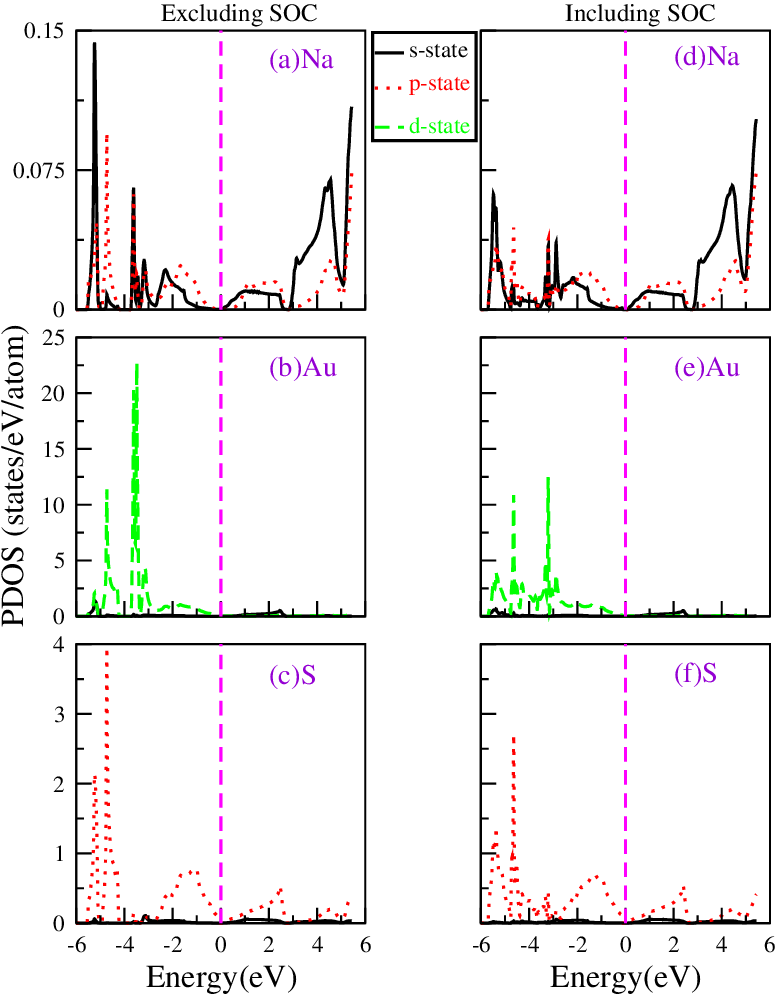}
\end{figure}

\begin{figure}
\caption{Electronic band structure for NaAuS with excluding spin-orbit coupling. (a) using LDA and PBEsol, (b) using PBE and revPBE. Zero energy corresponds to Fermi level.\\
\\}
\includegraphics[width=1.2\linewidth, height=8.5cm]{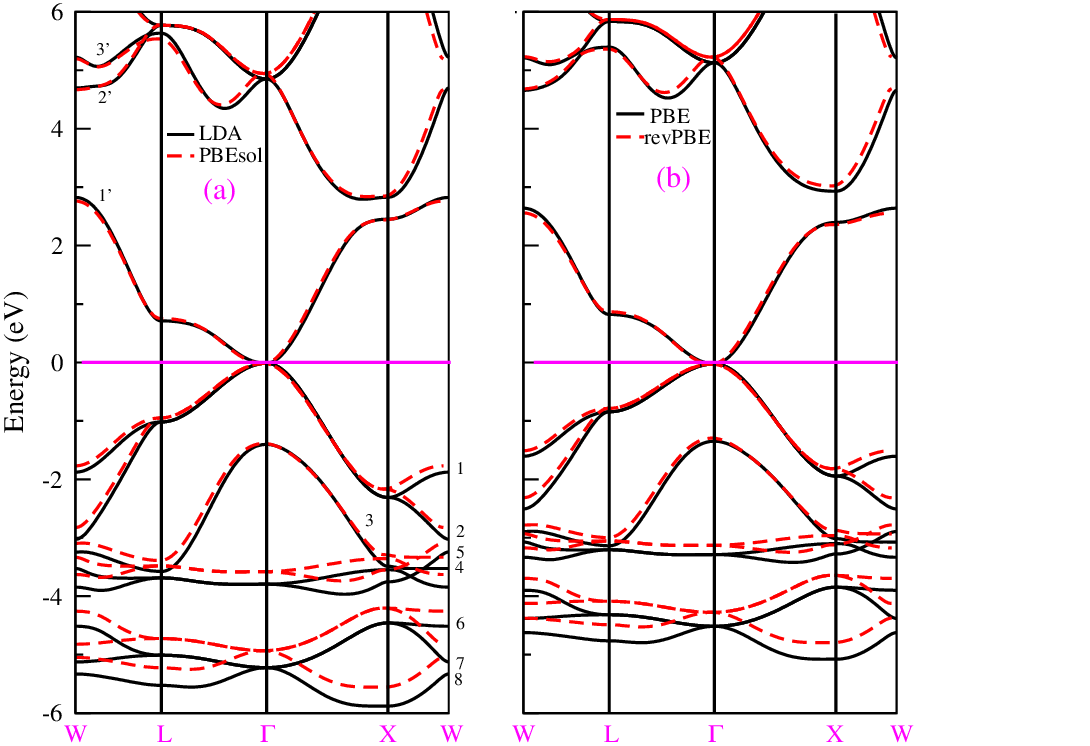}
\end{figure}

   The plot of partial density of states (PDOS) of Na, Au and S atoms of NaAuS with and without including SOC in the calculations only for PBEsol exchange correlation functional is shown in the Fig. 5(a-f). This is due to the fact that among LDA, PBEsol, PBE and revPBE exchange correlation functionals, PBEsol is newest one and show almost similar PDOS as compared to other exchange correlation functionals. Here, we discuss the PDOS of Na, Au and S atoms of NaAuS without including SOC in the calculations. Below Fermi level, the dominant electronic contribution comes from 3$s$  and 3$p$ states of Na atom, which are $\sim$45\% and $\sim$55\%, respectively. For Au atom, the most dominant electronic contribution comes from 5$d$ state ($\sim$93\%), while the small contribution comes from other states. For S atom, the dominant electronic character comes from 3$p$ state ($\sim$95\%) as compared to 3$s$ state. Now, we discuss the PDOS of Na, Au and S atoms above the Fermi level. The electronic contribution of 3$s$ and 3$p$ states to PDOS of Na atom are $\sim$66\% and $\sim$34\%, respectively. For Au atom, the contribution of 6$s$ and 5$d$ states are negligible to PDOS. The dominant electronic character for S atom comes from 3$p$ state ($\sim$75\%) as compared to 3$s$ state. Almost a similar electronic contribution from various states to PDOS of Na, Au and S atoms of NaAuS is observed by including the SOC in the calculations and is shown in Fig. 5(d-f).

\begin{figure}
\caption{Electronic band structure for NaAuS compound with including spin-orbit coupling using PBEsol.\\
\\}
\includegraphics[width=1.0\linewidth, height=8.5cm]{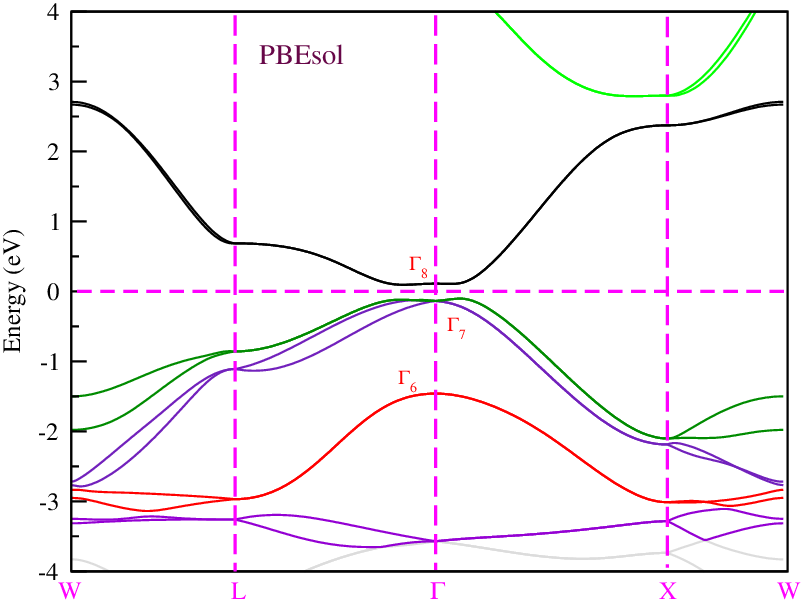}
\end{figure}
       
   The spin unpolarised dispersion curves along the high symmetry direction of first Brillouin zone obtained in calculations for NaAuS compound for above mentioned exchange correlation functionals is shown in Fig. 6. The high symmetric k-points for FCC structure are W, L, $\Gamma$ and X, respectively. Firstly, we discuss the band structure of NaAuS for LDA exchange correlation functional. Top of the energy band of VB and bottom of the energy band of conduction band (CB) touches at $\Gamma$ point. The first two energy bands of VB from Fermi level are doubly degenerate from $\Gamma$ to L and $\Gamma$ to X directions. Along L to W and X to W directions, the degeneracy of energy bands is completely lifted. $\Gamma$ to L direction, 4 and 5 energy bands of VB from Fermi level are doubly degenerate, while $\Gamma$ to X and L to W directions both energy bands are non degenerate. However, both energy bands are crossing to each other with 3 energy band almost middle of X and W points. 6, 7 and 8 energy bands are triply degenerate at $\Gamma$ point. Along $\Gamma$ to L and $\Gamma$ to X directions, only 6 and 7 energy bands remains degenerate. While, from L to W and X to W, the degeneracy of 6 and 7 energy bands is lifted. Now, we discuss the various energy bands from the Fermi level of CB. 2' and 3' energy bands of CB are degenerate at $\Gamma$ point. However, the degeneracy of these bands is lifted from $\Gamma$ to L and $\Gamma$ to X direction. Also, it is clear from the figure that almost similar behaviour of energy bands of both VB and CB is observed for PBEsol, PBE and revPBE exchange correlation functionals. However, energy bands of VB are shifted towards the Fermi level, when the exchange correlation functional changes from LDA$\rightarrow$PBEsol$\rightarrow$PBE$\rightarrow$revPBE. Near the Fermi level, all energy bands of VB of every exchange correlation functional are overlapped with each other. Almost similar behaviour of energy bands of VB and CB is observed for every exchange correlation functional with including SOC in calculations. Here, we discuss the band properties of NaAuS compound for PBEsol exchange correlation functional as it is the newest among all these exchange correlation functionals. It is clear from Fig. 7 that the every energy band of VB and CB are splitted into two states by including SOC. The calculated values of direct band gap at $\Gamma$ point is predicted by LDA (PBEsol) and PBE (revPBE) exchange correlation functionals are $\sim$0.26 eV ($\sim$0.25 eV) and $\sim$0.24 eV ($\sim$0.23 eV), respectively as shown in Table 1.
         
   At last, we discuss the band inversion property of NaAuS at $\Gamma$ point as it is the most common tool to identify the topological insulating behavior of a compound. Here, we focus on $\Gamma_6$ and $\Gamma_8$ point when SOC is included in calculations, which is shown in Fig. 7. After the detailed analysis of the bands indicate that $\Gamma_6$ point is rich with Au 6$s$ state and the contribution of S 3$p$ state is greater than Au 5$d$ state at the $\Gamma_8$ point when SOC is not included in calculations. However, when the effect of SOC is included, $\Gamma_6$ point does not change its own contributor, where the contribution of Au 5$d$ state is increased at $\Gamma_8$ point as compared to S 3$p$ state. From our basic knowledge, we know that outer electronic state will stay at higher energy with respect to inner state. Au 5$d$ is a inner state than Au 6$s$ state. But, after including SOC Au 6$s$ state goes to lower energy level than the Au 5$d$ state. Hence, Au $s-d$ band inversion between $\Gamma_6$ and $\Gamma_8$ point is observed here. The band inversion strength for LDA, PBEsol, PBE and revPBE exchange correlation functionals are calculated by using relation, ${\Delta=[E_{\Gamma_8}-E_{\Gamma_6}]}$ \cite{Al}. The calculated values of band inversion strength (shown in Table 1) for LDA, PBEsol, PBE and revPBE exchange correlation functionals are $\sim$1.58, $\sim$1.57, $\sim$1.50 and $\sim$1.47 eV, respectively.

\section{Conclusions} 
    
    The detail electronic structure study of half-Heusler NaAuS compound has not been done theoretically. Here, we have studied the various comparative physical properties such as, structural and electronic for NaAuS by using LDA, PBEsol, PBE and revPBE exchange correlation functionals. The calculated values of equilibrium lattice constant (bulk modulus) for LDA, PBEsol, PBE and revPBE exchange correlation functionals are found to be $\sim$6.128 $\AA$ ($\sim$66.6 GPa), $\sim$6.219 $\AA$ ($\sim$56.4 GPa), $\sim$6.353 $\AA$ ($\sim$46.5 GPa) and $\sim$6.442 $\AA$ ($\sim$39.3 GPa), respectively. Hence, the order of calculated values of bulk modulus for these exchange correlation functionals are consistent with the order of calculated values of equilibrium lattice parameters because of the inverse relationship between bulk modulus and lattice parameter. Among these functionals, the total density of states below the Fermi level were found to be spread in more region for LDA and less region for revPBE, which was also found to be consistent with the order of bulk modulus for these exchange correlation functionals. In presence of spin-orbit coupling, a direct band gap was observed for NaAuS compound, which was found to be $\sim$0.26, $\sim$0.25, $\sim$0.24 and $\sim$0.23 eV for LDA, PBEsol, PBE and revPBE exchange correlation functionals, respectively. The band inversion was observed at $\Gamma$ point, which indicates that NaAuS compound shows topological insulating behaviour. The calculated values of band inversion strength for LDA (PBEsol) and PBE (revPBE) exchange correlation functionals were found to be $\sim$1.58 eV ($\sim$1.57 eV) and $\sim$1.50 eV ($\sim$1.47 eV), respectively.
 
\section{Acknowledgements}

  S.L. is thankful to UGC, India, for financial support.

\end{document}